\documentclass[twocolumn,secnumarabic,showpacs,amsmath,amssymb, nobibnotes, aps, pra]{revtex4-1}

\usepackage{graphicx}

\begin{document}

\title{Analytical solution of the Schr\"{o}dinger equation for the hydrogen molecular ion $H_2^+$}
\author{Alexander V. Mitin}
\email[]{mitin@phys.chem.msu.ru}
\affiliation{The Chemistry Department of the Moscow State University, 119991 Moscow, Russia \\
$^1$Moscow Institute of Physics and Technology,  9 Institutskiy per., Dolgoprudny, Moscow Region, 141700, Russia
}
\date{\today}
%\date{August 10, 2010}%
\begin{abstract}
The full analytical solution of the Schr\"{o}dinger equation for the hydrogen molecular ion $H_2^+$ (special case of the quantum tree-body problem with the Coulomb interaction) is obtained first. The solution shows that the total wave function is a two-term function in the sense that it is a linear combination of the two linear independent wave functions. The two-term character of the total wave function was visualized in calculations of the total electron density of $H_2^+$ at different internuclear separations. 

\end{abstract}
\pacs{31.10.+z}
\maketitle
%\tableofcontents

The hydrogen molecular ion $H_2^+$ is a very important case of the quantum tree-body problem with the Coulomb interaction. It is well known that the variables of the Schr\"{o}dinger equation for the $H_2^+$ molecular ion in the Born-Oppenheimer approximation are separated in elliptic coordinates $\eta=(r_1-r_2)/R$, $\xi=(r_1+r_2)/R$, where $R$ is the internuclear distance and $r_1$, $r_2$ are distances between the electron and nuclei 1 and 2, correspondingly \cite{Burrau_1927}. In this case, the wave function for the $\Sigma$ ground state can be presented as 
$$
\Psi=X(\xi,R)Y(\eta,R), 
$$
where the functions $X(\xi,R)$ and $Y(\eta,R)$ satisfy the following equations:
\begin{eqnarray}
\begin{gathered}
\left(\xi^2-1\right)\frac{d^2X}{d\xi^2}+2\xi\frac{dX}{d\xi}+ \\ \left(\frac{1}{2}ER^2\xi^2-2R\xi+A\right)X=0 \label{lable_01}
\end{gathered}
\end{eqnarray}
\begin{eqnarray}
\left(1-\eta^2\right)\frac{d^2Y}{d\eta^2}-2\eta\frac{dY}{d\eta}-
\left(\frac{1}{2}ER^2\eta^2+A\right)Y=0, \label{lable_02}
\end{eqnarray}
where $E$ and $A$ are energy and separation constant which have to be determined. Equation (\ref{lable_02}) is most important among these two equations because its solution is mainly defined the property of the molecular wave function along the internuclear axis. 

An analytic solution for the Eq. (\ref{lable_01}) was given in \cite{Jaffe_1934}, while an analytic solution of Eq. (\ref{lable_02}) was unknown up to present study. However, its solution can be obtained by noting that $Y \to e^{\pm \sqrt{A}\eta}$ when $\eta \to 0$. Then, the two linear independent solutions of this equation can be obtained by substituting two anzatz functions into it
$$
Y_1=g_1(\eta)e^{+\sqrt{A}\eta} ,
$$
$$
Y_2=g_2(\eta)e^{-\sqrt{A}\eta} ,
$$
where $g_1(\eta)$ and $g_2(\eta)$ are new functions. This gives us two equations for $g_1(\eta)$ and $g_2(\eta)$
$$
\begin{gathered}
\left(1-\eta^2\right)\frac{d^2g_1}{d\eta^2}+\left[2\sqrt{A}\left(1-\eta^2\right)-2\eta\right]\frac{dg_1}{d\eta}+ \\
\left[A\left(1-\eta^2\right)-2\sqrt{A}\eta-\left(\frac{1}{2}ER^2\eta^2+A\right)\right]g_1=0 , 
\end{gathered}
$$
$$
\begin{gathered}
\left(1-\eta^2\right)\frac{d^2g_2}{d\eta^2}-\left[2\sqrt{A}\left(1-\eta^2\right)+2\eta\right]\frac{dg_2}{d\eta}+ \\
\left[A\left(1-\eta^2\right)+2\sqrt{A}\eta-\left(\frac{1}{2}ER^2\eta^2+A\right)\right]g_2=0 . 
\end{gathered}
$$
Their polynomial solutions can be found by applying the usual method used for the $H_2^+$ problem \cite{MorseStueckelberg_1929, Teller_1930}.

The first most important question is algebraic structure of the general solution of differential Eq. (\ref{lable_02}). It is written as the linear combination of two linear independent solutions $Y_1$ and $Y_2$:
$$
Y=C_1(\eta)e^{+\sqrt{A}\eta}g_1(\eta)+C_2(\eta)e^{-\sqrt{A}\eta}g_2(\eta) . 
$$
This solution is symmetric with respect to the sign change of $\eta$. Therefore, it can be rewritten in the form
\begin{eqnarray}
\begin{gathered}
Y=D_1(abs(\eta))e^{+\sqrt{A}abs(\eta)}f_1(\eta)+ \\
  D_2(-abs(\eta))e^{-\sqrt{A}abs(\eta)}f_2(\eta) , \label{lable_03}
\end{gathered}
\end{eqnarray}
which explicitly displays physical meaning of both terms. The first term in (\ref{lable_03}) can be interpreted as the molecular wave function of cylindrical symmetry, because in interval $[-1,1]$ it has a minimum at $\eta=0$ and maxima at $\eta=\pm1$, while the second one can be interpreted as the quasi-atomic wave function of spherical symmetry, because it has a maximum at $\eta=0$.  

The algebraic structure of the wave function $Y$ (\ref{lable_03}) directly shows the correlation of the total wave function with the wave function of the united atom $He^+$ studied first in \cite{Bethe_1933} and in many other investigations including detailed consideration given in \cite{KPS_1976}. Taking this into account we will call  the total wave function a two-term total wave function because it is a linear combination of two linear independent wave functions.

\begin{figure*}
\includegraphics[scale=0.7]{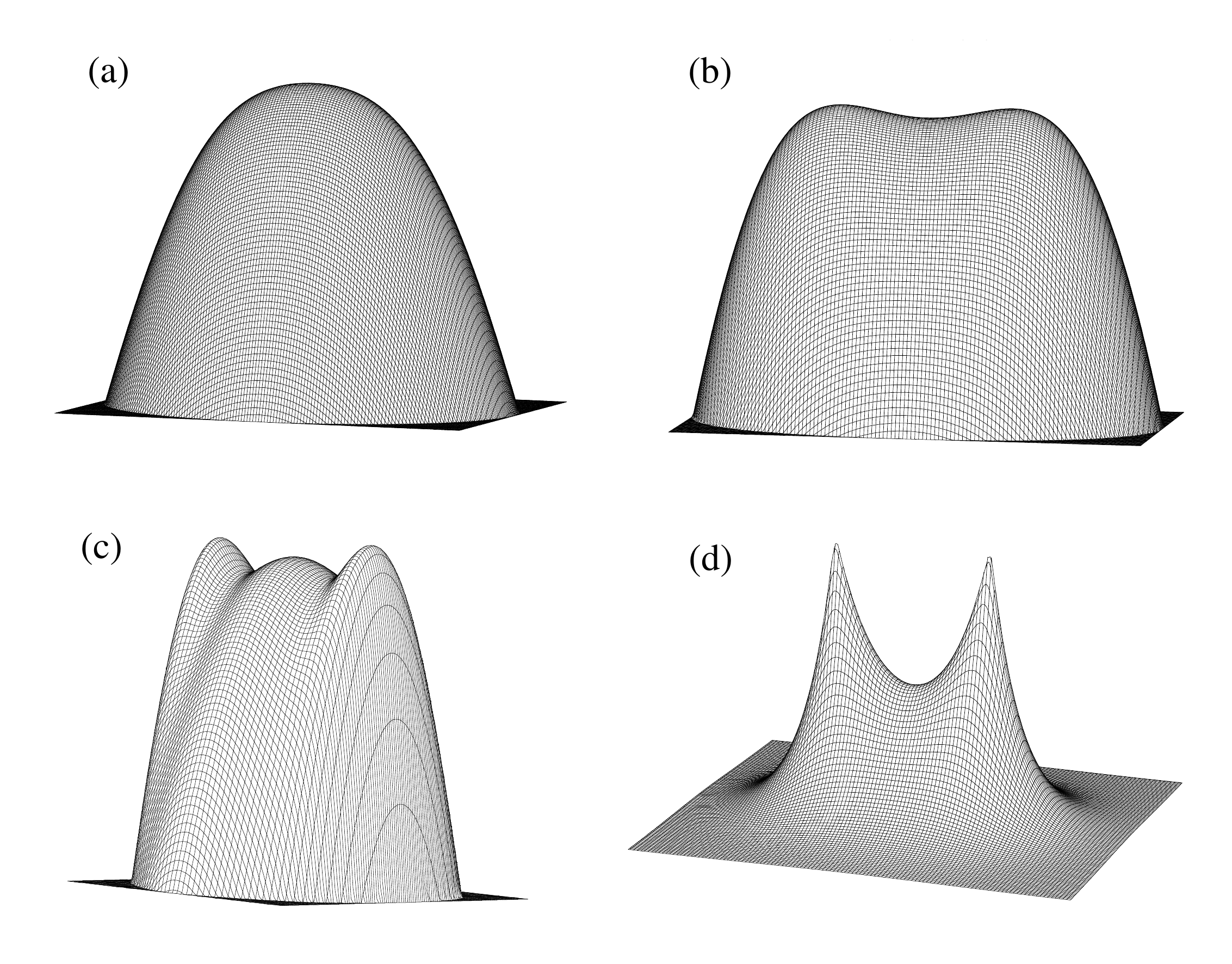}
\caption{Electron densities in $H_2^+$ at different internuclear distances (in a.u.): (a) - 0.008, (b) - 0.012, (c) - 0.025, (d) - 2.0.}
\label{fig_01}
\end{figure*}

It is important to emphasize here that the algebraic structure of the total molecular wave function and its asymptotic or correlation property in the direction of the united atom does not contradict the two-body cusp condition for many-body wave functions \cite{Kato_1957}. It is sufficient to note that the two-body cusp condition is correct only in the full coordinate space, which has a dimension of $3N$, where $N$ is the number of considered particles \cite{Kato_1957}. The full coordinate space is a non-physical space. In three-dimensional physical space the total molecular wave function has the asymptotic or correlation properties but no two-body cusp conditions. 

Numerical calculations of three-dimensional electron density distributions in $H_2^+$ at different internuclear distances give additional confirmation that the total wave function is a two-term function.

In this connection, the total electron density of the $H_2^+$ was calculated by the Hartree-Fock method using the Gaussian 94 program \cite{G_1994}. The calculations were performed with the nuclei centered $15s6p5d4f3g2h1i$ Gaussian basis set. Obtained total electron density distributions in $H_2^+$ are presented on Fig. \ref{fig_01}.

From the fact that the total wave function is a two-term function follows that at short internuclear separations, where the quasi-atomic wave function is dominant, the total electron density distribution should be in the form of a perturbed spheroid. Explicitly, such a form of the total electron density  distribution is clearly observed on Fig. \ref{fig_01}a at internuclear separation $R=0.008$ a.u. The intermediate region begins at larger $R$, where the amplitude of the molecular wave function becomes noticeable. This leads to the appearance of two maxima on the total electron density distribution. Explicitly such a distribution with two maxima at the positions of two nuclei can be observed on Fig. \ref{fig_01}b at $R=0.012$ a.u. This distance is in excellent agreement with the investigation of the Bethe perturbation theory \cite{Bethe_1933} in \cite{Wind_1965}, where it was shown that noticeable deviations of this theory from the exact solution for $H_2^+$ begin at this distance. In the intermediate region the amplitudes of both wave functions have to become comparable at some $R$ and, hence, electron densities from both wave functions have to be observable on the total electron density distribution. Namely, this case is presented on Fig. \ref{fig_01}c at $R=0.025$ a.u. The electron density from the quasi-atomic wave function of spherical symmetry is located in the middle between two nuclei and the electron density from the molecular wave function of cylindrical symmetry is well recognizable by the two maxima at the positions of the two nuclei. Finally, at large $R$ the molecular wave function has to become dominant and the distribution of total electron density has to have the usual form. Such a distribution of the total electron density can be seen on Fig. \ref{fig_01}d at $R=2.0$ a.u.

Thus, the analytical solution of the Schr\"{o}dinger equation for the hydrogen molecular ion $H_2^+$, and the total electron density distributions presented above show that the total wave function is a linear combination of two-linear independent wave functions (a two-term function) in which the amplitudes of wave functions change with the variation of the internuclear separation. Near the equilibrium point the molecular wave function of the total wave function is dominant, while at short distances the quasi-atomic wave function becomes dominant.

\clearpage
\textbf{List of the figure captions}.

\textbf{Figure 1}. Electron densities in $H_2^+$ at different internuclear distances (in a.u.): (a) - 0.008, (b) - 0.012, (c) - 0.025, (d) - 2.0.

\end{document}